\documentclass[english,aps,twocolumn,groupedaddress,pra,superscriptaddress,amsmath,amssymb,noeprint,nolongbibliography]{revtex4-2}

\usepackage[T1]{fontenc}
\usepackage{mathrsfs}
\usepackage{amsmath, amstext}
\usepackage{amssymb}
\usepackage{graphicx}
\usepackage{esint}
\usepackage{physics, braket}
\usepackage{xcolor,soul}
\usepackage{url}



\usepackage{babel}
\begin{document}
\title{Time-dependent Rabi frequencies to protect quantum operations on an atomic qutrit by continuous dynamical decoupling}
\author{Adonai Hil\'ario da Silva}
\email{adonai.silva@usp.br}
\affiliation{S\~ao Carlos Institute of Physics, University of S\~ao Paulo, PO Box 369,
13560-970, S\~ao Carlos, SP, Brazil}
\author{Reginaldo de Jesus Napolitano}
\affiliation{S\~ao Carlos Institute of Physics, University of S\~ao Paulo, PO Box 369,
13560-970, S\~ao Carlos, SP, Brazil}
\author{Felipe Fernandes Fanchini}
\affiliation{Faculdade de Ci\^encias, UNESP - Universidade Estadual Paulista, 17033-360,
Bauru, SP, Brazil}
\affiliation{QuaTI - Quantum Technology \& Information, 13560-161, São Carlos, SP, Brazil}
\author{Bruno Bellomo}
\affiliation{Universit\'{e} de Franche-Comt\'{e}, Institut UTINAM, CNRS UMR 6213, Observatoire des Sciences de l'Univers THETA, 41 bis avenue de l'Observatoire, F-25010 Besan\c{c}on, France}

\begin{abstract}
We investigate the form required for the time-dependent Rabi frequencies involved in a procedure capable to protect the action of quantum gates on an atomic qutrit by means of external fields continuously decoupling the system from the environmental noise.
Several simulations are considered to protect the action of quantum-gate models, including randomly chosen ones. We argue that the requirements for the Rabi frequencies could be nowadays  experimentally met. We also investigate the transition from one gate operation to another, including protecting a qutrit memory state. We finally apply our methodology to protect from noise the application of an algorithm capable of distinguishing the parity of permutations of three elements.
\end{abstract}

\maketitle

\section{Introduction}

Quantum coherence is the valuable resource that physically distinguishes
between quantum and classical computers. Thus, preserving coherence
or recovering it from decoherence is fundamental for high-fidelity processing
of quantum information. The first specific paper about dynamically
suppressing decoherence of a quantum state involved a single qubit~\cite{PhysRevA.58.2733}.
However, at least two qubits are necessary for a single entangling
gate~\cite{michaelnielsen2011} and useful quantum-information processing
requires more than only a few universal quantum gates~\cite{brylinski2002universal}.
As quantum algorithms become more sophisticated, the complexity of
the quantum circuits for running them inevitably increases, as does
the dimension of the required Hilbert spaces. Recent advances have
introduced photonic qudit systems~\cite{PhysRevLett.94.100501,Reimer1176,Kues2017},
with their quantum states belonging to high-dimensional Hilbert spaces.
As an impressive example of how high a dimension recent technology
can reach, we cite the realization of two entangled photonic qudits
in 100 dimensions~\cite{Kues2017}. As examples of useful applications
of quantum-information processing using qudits we mention quantum-key
distribution~\cite{PhysRevLett.85.3313,PhysRevLett.88.127901,PhysRevLett.88.127902,PhysRevA.67.012310,PhysRevLett.98.060503},
quantum computation~\cite{PhysRevA.66.022317,PhysRevLett.94.230502,PhysRevA.87.022341,PhysRevA.95.052317,PhysRevLett.120.160502},
and even basic investigations on the nature of quantum mechanics~\cite{PhysRevLett.88.040404}.
Therefore, approaches to avoid or recover from errors introduced by
decoherence are bound to attract ever-increasing attention~\cite{doi:10.1142/S0129054103002011,PhysRevX.2.041021,PhysRevLett.113.230501,PhysRevA.91.042331,PhysRevA.104.L060401}.
Evidence has been accumulating that such higher-dimensional states
and the corresponding quantum gates acting on them provide additional
support against decoherence, aside from improving the efficiency of quantum
circuits~\cite{PhysRevA.75.022313,Lanyon2009,PhysRevA.84.052313}.
Proposals to implement qudit devices abound, not only using optical
systems, but also superconducting~\cite{PhysRevA.84.052313,PhysRevLett.105.050501}
and atomic-spin~\cite{PhysRevA.85.022302,PhysRevLett.99.163002} devices.

We have recently presented a generalized procedure for continuously
and dynamically decoupling an ensemble of qudits from general environmental
perturbations~\cite{napolitano2019protecting}. Using generalized continuous
dynamical decoupling (GCDD) we can protect any qudit system
from general noise even during the operation of an arbitrary qudit
gate. In Ref.~\cite{napolitano2019protecting} we illustrate a practical
implementation of the GCDD method with a very detailed qutrit model,
consisting of the three magnetic hyperfine states of the ground energy
level of $^{87}\mathrm{Rb},$ during the action of a qutrit Hadamard
gate. The quantum gate and the control are implemented using laser
beams whose intensities and phases are modulated according to our
prescription. Our model relies on the control over the Rabi frequencies
of nine distinguishable laser beams, in an orchestrated scheme that, in principle, would allow us to generate the dynamical decoupling from noise, simultaneously executing the intended one-qudit gate operation. However, this situation raises the question whether
such prescribed modulation of the control fields could be implemented
using current technology, since the required time-varying Rabi frequencies
could prove themselves unrealizable.

Here we explicitly study the form of the time-dependent Rabi frequencies required to decouple an atomic
qutrit from the environmental noise while also executing
quantum operations, also validating the approximations involved in the protection scheme. Most of the
previous investigations used pulsed dynamical decoupling of qubits~\cite{PhysRevLett.82.2417,PhysRevLett.83.4888,doi:10.1080/09500340408235288,PhysRevA.85.032331}.
In the case of laser fields driving two-photon transitions in qutrits,
Gaussian-shaped time-dependent Rabi frequencies~\cite{PhysRevA.85.032331}
have been studied in detail. The GCDD method does not use pulses,
but in the case of qudits ($d$-level systems)~\cite{napolitano2019protecting}, it prescribes
time-dependent Hamiltonians to generate any $\mathrm{SU}(d)$
operation continuously. Here we consider our illustrative model for an
atomic qutrit, addressing the possibility to tailor the necessary
Rabi frequencies using currently available technological capabilities.
We estimate the validity of approximations and suggest values of the parameters that are currently
achievable. We simulate the protection scheme under quite general environmental
perturbations, during the operation of different kinds of quantum gates that act on
the qutrit. We make use of continuous external fields, whose period $t_{0}$ is assumed
shorter than the correlation time $\tau_c$ of the noise, to emulate an appropriate
periodic time-dependent unitary evolution. It is worth emphasizing that, although here we address the qutrit model using a laser-atom system, the same analysis could be extended to qudit systems involving higher dimensions, as for example, another set of hyperfine levels of an atomic system, or to a treatment of more than a single qudit. We do not expect any qualitatively different impediments in such an extension to the general conclusions about our analysis presented here. We also stress that even if we use a specific atomic implementation for the qutrit, we consider as source of the external noise, in the same spirit of Ref.~\cite{napolitano2019protecting}, a very general model for the environment which is not specifically connected to a given experimental platform.

This article is organized as follows. Section~\ref{Sec:GCDD} briefly summarizes the GCDD method. 
In Sec.~\ref{Sec:qutrit} we review our qutrit model  
and we give the explicit expression of the external
fields required for implementing the required GCDD Hamiltonian. In Sec.~\ref{Sec:Rabi} we study
the required Rabi frequencies for various one-qutrit quantum operations,
we verify the validity of our approximations, and we argue that the requirements for the Rabi frequencies
could be currently met. We also show that the GCDD procedure can be used to protect the execution of a quite simple quantum algorithm \cite{gedik2015computational}, where a single qutrit is used to determine the parity of the permutations of a set of three objects. Finally, Sec.~\ref{Sec:Conclusions} presents our conclusions. 

\section{A summary of the GCDD procedure}\label{Sec:GCDD}

The GCDD prescription detailed in Ref.~\cite{napolitano2019protecting}, is a generalization of the continuous
dynamical decoupling of qubits~\cite{PhysRevA.75.022329,PhysRevA.76.032319,PhysRevA.76.062306,PhysRevA.91.042325,Yalcinkaya2019}
and consists of a continuously-varying control Hamiltonian and a modification
of the intended quantum operation that, together, aim to suppress the negative effects of the environmental noise. Indeed, the final state results to be closer to the state which would have been obtained if the ideal noise-free version of the intended operation had taken place, with respect to the case when the operation is performed in the presence of the noise but without protection. The GCDD procedure can be
extended to the case of an arbitrary number of qudits, identical
or not~\cite{napolitano2019protecting}. In this section, we briefly summarize our previous work regarding the GCDD prescription, by focusing on the case of a single qudit. The main point concerns the introduction of a basic unitary evolution that satisfies a weaker condition of the general dynamical decoupling condition~\cite{PhysRevA.71.022302}, which is sufficient to effectively decouple the system from its environment~\cite{napolitano2019protecting}. 

According to the GCDD prescription, the total Hamiltonian is written as
\begin{equation}
H_{\mathrm{tot}}(t) = \left[H_{c}(t) +H_{\mathrm{gate}}(t)\right]\otimes \mathbb{I}_E +\mathbb{I}_d \otimes  H_{E}+H_{\mathrm{int}},\label{eq:Htot}
\end{equation}
where $\mathbb{I}_E$ and $\mathbb{I}_d$ are the identity operators of, respectively, the environmental and  the qudit Hilbert spaces.
Here $H_{\mathrm{gate}}(t)$ is the Hamiltonian associated to the ideal gate Hamiltonian $H_{G}$ one wants to apply to the qudit. The quantum-gate operation starts at $t=0$ and stops at $t=\tau$. According to our protection scheme, the Hamiltonian $H_{\mathrm{gate}}(t)$ is defined by
\begin{equation}
H_{\mathrm{gate}}(t) \equiv  U_{c}(t)H_{G}U_{c}^{\dagger}(t),\label{Hgate}
\end{equation}
where $U_{c}(t)$ is the control unitary transformation associated to external control fields whose action is described by $H_{c}(t)$.
$H_{E}$ is the free Hamiltonian
of the environment, and $H_{\mathrm{int}}$
describes the interaction between the qudit and the environment. The noise effects are
assumed to be resulting from the interaction between the qudit and the environment, described by the very general Hamiltonian
\begin{eqnarray}
H_{\mathrm{int}} & = & \sum_{r=0}^{d-1}\sum_{s=0}^{d-1}\ketbra{r}{s}\otimes B_{rs},\label{eq:Hint}
\end{eqnarray}
where $B_{rs}$, for $r,s=0,1,\dots,d-1$, are operators that
act on the environmental states and $\ket{k}$, for
$k=0,1,\dots,d-1$, are $d$ normalized states composing a logical basis set for the qudit Hilbert space of dimension $d$.

We remark that the free Hamiltonian of the qudit can be thought as included in $H_G$ or assumed to be eliminated before the application of the GCDD scheme. The protection protocol is then applied to a system which, in the absence of $H_G$, is degenerate since all qudit levels have the same energy.

The procedure also involves the Hermitian operator  $H_{L}$ whose action on the logical basis states
is given by
\begin{equation}
H_{L}\ket{k}   \equiv  k\hbar\omega_{d}\ket{k} ,\label{eq:HL}
\end{equation}
for $k=0,1,\dots,d-1$, where $ \omega_{d} \equiv \omega_{0}d $ and $\omega_{0}$ is the control frequency corresponding to the dynamical-decoupling period $t_0 \equiv 2\pi/\omega_{0}$.
We also make use of the Hermitian operator $H_{F}$
whose action on the quantum-Fourier transformed basis is defined as
\begin{equation}
H_{F}\ket{\psi_n} \equiv n\hbar\omega_{0}\ket{\psi_n} ,\label{eq:HF}
\end{equation}
where
\begin{equation}
\ket{\psi_n} \equiv \frac{1}{\sqrt{d}}\sum_{j=0}^{d-1}\exp(\frac{2\pi i}{d}jn)\ket{j} ,\label{eq:psin}
\end{equation}
for $n=0,1,\dots,d-1.$ The control unitary transformation is then given in terms of $H_{L}$ and $H_{F}$ as
\begin{equation}
U_{c}(t)\equiv  \exp(-i\omega_{r}t) \exp(-i\frac{H_{L}}{\hbar}t) \exp(-i\frac{H_{F}}{\hbar}t),\label{eq:Uc(t)}
\end{equation}
where, for convenience, we have introduced a real constant $\omega_{r}$ as
\begin{equation}
\omega_{r}\equiv  -\frac{\mathrm{Tr}\left(H_{L}\right)+\mathrm{Tr}\left(H_{F}\right)}{\hbar d}.\label{eq:varphi}
\end{equation}
We stress that the operator $U_{c}(t)$ is built to be periodic and its period is $t_0$. Using
the definition of $U_{c}(t)$ in Eq.~\eqref{eq:Uc(t)},
we can calculate the control Hamiltonian as 
\begin{eqnarray}
H_{c}(t) & = & i\hbar\frac{\mathrm{d}U_{c}(t)}{\mathrm{d}t}U_{c}^{\dagger}(t)\nonumber \\
 & = &  \hbar \omega_r \mathbb{I}_d+ H_{L}+\exp(-i\frac{H_{L}}{\hbar}t)H_{F}\exp(i\frac{H_{L}}{\hbar}t).\quad\:\label{eq:Hc(t)}
\end{eqnarray}

It is now possible to show that the following  sufficient condition for dynamically decoupling \cite{PhysRevA.71.022302} for the interaction defined by Eq.~\eqref{eq:Hint} is satisfied:
\begin{equation}
\int_{0}^{t_{0}}\mathrm{d}t\,\left[U_{c}^{\dagger}(t) \otimes \mathbb{I}_E\right] H_{\mathrm{int}} \left[ U_{c}(t)\otimes \mathbb{I}_E \right]=  \frac{t_0}{d} \mathbb{I}_d \otimes \sum_{r=0}^{d-1} B_{rr}.\label{eq:dd}
\end{equation}
Here the period of $U_c(t)$, $t_0$, is to be chosen so that it is short enough compared with the bath correlation time $\tau_c$, and so that the gate time $\tau$ is an integer multiple of $t_{0}$.

In the laboratory we need to generate
external fields such that they interact with the qudit according to the Hamiltonian $H_{\mathrm{lab}}(t)=H_{c}(t)+H_{\mathrm{gate}}(t)$. We assume that the ideal gate Hamiltonian $H_{G}$
can be expanded in the computational basis as
\begin{equation}
H_{G}  =  \hbar\sum_{r=0}^{d-1}\sum_{s=0}^{d-1}g_{r,s}\ketbra{r}{s},\label{Hd}
\end{equation}
with $g_{s,r}^{\ast} = g_{r,s}$ since $H_{G}$ is Hermitian. As explained in Ref.~\cite{napolitano2019protecting},
we have taken $H_{\mathrm{gate}}(t)$ as defined in Eq.~\eqref{Hgate}
so that, in the presence of the control fields, the intended quantum operation can be realized even when there is noise with high fidelity, without the control fields also decoupling the gate action. We define
$\hbar G = \hbar g_{0}\mathbb{I}_{d}-H_{G}$, where $g_{0}$ is the largest eigenvalue of $H_{G},$ so that $G$
is positive semi-definite. We also define the following semi-positive operators: $H_{L}^{\prime} \equiv \hbar\left(d-1\right)\omega_{d}\mathbb{I}_{d}-H_{L}$ and $H_{F}^{\prime} \equiv \hbar\left(d-1\right)\omega_{0}\mathbb{I}_{d}-H_{F}$. We thus obtain
\begin{equation}
H_{\mathrm{lab}}(t)  =  \hbar\omega_{l}\mathbb{I}_{d}-\hbar\Upsilon(t)\Upsilon(t),\label{Hlab}
\end{equation}
with $\omega_{l} \equiv g_{0}+\omega_{r}+\left(d^{2}-1\right)\omega_{0}$
and 
\begin{equation}
\Upsilon(t) \equiv   \sqrt{ \frac{V}{\hbar}},\label{Upsilon}
\end{equation}
where
\begin{equation}
V\!= \! H_{L}^{\prime}+\exp\!\left(-i\frac{H_{L}}{\hbar}t\right)\!H_{F}^{\prime}\exp\!\left(i\frac{H_{L}}{\hbar}t\right)\!+\hbar U_{c}(t)GU_{c}^{\dagger}(t),
\end{equation}
which is a well-defined square root, since $V$
is a positive semi-definite operator. 

As one can deduce from Eq.~\eqref{eq:Hc(t)}, the protection procedure does not depend on the characteristics of the environment and on the coupling constants regulating the interaction between the system and the environment. However, the effectiveness of the protection depends on the ratio between the bath correlation time and the control period.

\section{An atomic qutrit}\label{Sec:qutrit}

\begin{figure}[t!]
\begin{centering}
		\includegraphics[scale=0.225]{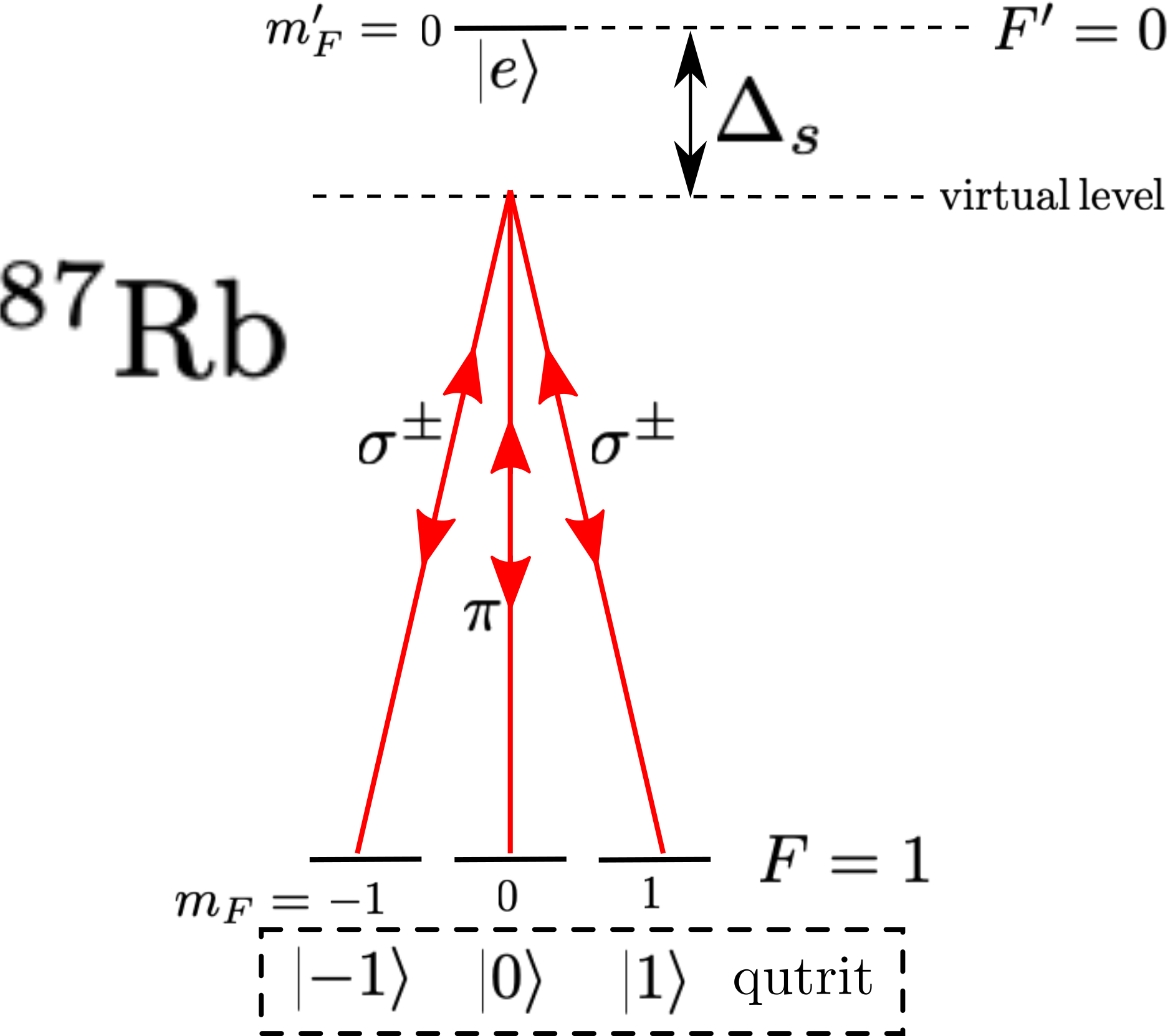}
		\par\end{centering}
	\caption{\label{Fig1}The hyperfine states of the $\mathrm{D}_{2}$ transition
		of $^{87}\mathrm{Rb}$ that we need in the present model (not to scale). We use two-photon transitions
		with three different detunings $\Delta_{s}$, for $s=1,2,3,$ and,
		for each of these detunings, we use $\sigma^{\pm}$- and $\pi$-polarized
		laser light, as described in full detail in Ref.~\cite{napolitano2019protecting}.}
\end{figure}

Now we briefly review the qutrit model we have introduced in Ref.~\cite{napolitano2019protecting}, which may be consulted for the mathematical
derivations and other details. The qutrit comprises the  hyperfine ground manifold of magnetic states of a $^{87}\mathrm{Rb}$ atom, as shown in Fig.~\ref{Fig1}. These three ground states are degenerate in the absence of magnetic fields and
we denote them by $\ket{m}$, for $m=-1,0,1$. We name $\hbar \omega_g$ the energy of the ground states and $\hbar \omega_e$ the energy of the excited state $\ket{e}$.

The effective control fields involved in the GCDD procedure consist of a set of polarized laser beams.
We only use photon frequencies $\omega_s$ that are red detuned (this means that
the detunings $\Delta_{s}$, defined as $\Delta_{s}\equiv \omega_s-\omega_e+\omega_g$, are negative), from
the $F=1\leftrightarrow F^{\prime}=0$ transition. Thus, we can approximate
the relevant set of atomic states as $\left\{\ket{m}\text{, for } m=-1,0,1\right\}$, together with the excited
state $\ket{e}$, forming a restricted Hilbert
space that we denote by $\mathscr{H}_{4}$.
If the photons are detuned far enough to the red of the transitions
$\ket{m} \leftrightarrow \ket{e}$, for
$m=-1,0,1$ (virtual transitions), then the excited state is not going to be effectively populated, avoiding spurious transitions to the $F=2$ ground states
through spontaneous emission from $\ket{e}$. Accordingly, the effective qutrit consists of the states $\ket{m}$, with $m=-1,0,1$, whose Hilbert space is denoted by $\mathscr{H}_{3}$. The GCDD control over the states in $\mathscr{H}_{3}$ is accomplished through two-photon transitions by means of nine laser beams as comprehensively explained in 
 Ref.~\cite{napolitano2019protecting}. We make use of frequency and intensity modulation and we choose the frequencies and two-photon transitions so that a suitable
 form of the rotating-wave approximation is valid. The three states $\left\{\ket{m}\text{, for }m=-1,0,1\right\}$ form a qutrit computation basis. In the following we use the notation $\left\{\ket{m}\text{, for } m=-1,0,1\right\}$ instead of explicitly associating them to computational states $\left\{\ket{d}\text{, for } d=0,1,2\right\}$.

Following these parameter guidelines, now we must follow the prescription 
laid out in Ref.~\cite{napolitano2019protecting} to connect nine independent time-varying Rabi
frequencies $\Omega_{s,m }(t)$ ($s=1,2,3$ and $m=-1,0,1$), associated to the various physical transitions involved in the dynamics, with the Hamiltonians $H_{\mathrm{gate}}(t)$ and $H_{c}(t)$
of Eq.~\eqref{eq:Htot}. In particular the Rabi frequencies are defined as $\Omega_{s,m }(t)= \left(-1\right)^{q}D^{\ast}\mathscr{E}_{s,q}(t)/\hbar $, where $D$ is linked to the atomic electric-dipole operator $\mathbf{d}$, and $\mathscr{E}_{s,q}(t)$ are the amplitudes of nine external control laser beams. In particular, it holds $ \left\langle m\right|\mathbf{d}\left|e\right\rangle   =  D\boldsymbol{\hat{\varepsilon}}_{m}^{\ast}$, where $\boldsymbol{\hat{\varepsilon}}_{m}$ are polarization versors defined as 
$\boldsymbol{\hat{\varepsilon}}_{\pm1}  \equiv  \mp(\mathbf{\hat{x}}\pm i\mathbf{\hat{y}})/\sqrt{2}$,
representing, respectively, the $\sigma^{\pm}$ polarizations, and
$\boldsymbol{\hat{\varepsilon}}_{0}  \equiv  \mathbf{\hat{z}}$,
representing the $\pi$ polarization. Concerning the nine laser beams, the electric-field vectors can be written as
\begin{eqnarray}
&&\!\!\!\!\!\!\!\!\mathbf{E}_{\pm1}\left(t\right)  =  \sum_{s=1}^{3}\left[\mathscr{E}_{s,\pm1}\left(t\right)\boldsymbol{\hat{\varepsilon}}_{\pm1}\exp\left(-i\omega_{s}t\right) \right. \nonumber \\
& & \left.\qquad \quad + \mathscr{E}_{s,\pm1}^{\ast}\left(t\right)\boldsymbol{\hat{\varepsilon}}_{\pm1}^{\ast}\exp\left(i\omega_{s}t\right)\right],
\label{Epm}\nonumber \\
&&\!\!\!\!\!\!\!\!\mathbf{E}_{0}\left(t\right)  =  \boldsymbol{\hat{\varepsilon}}_{0}\sum_{s=1}^{3}\left[\mathscr{E}_{s,0}\left(t\right)\exp\left(-i\omega_{s}t\right)+\mathscr{E}_{s,0}^{\ast}\left(t\right)\exp\left(i\omega_{s}t\right)\right].\nonumber \\ 
\label{E0}
\end{eqnarray}
 In Ref.~\cite{napolitano2019protecting} we have shown that the Rabi
frequencies we need to generate the dynamics evolving according to a Hamiltonian equivalent to that of Eq.~\eqref{Hlab} up to a term proportional to the identity, which only introduces an immaterial global phase factor (this aspect is not even present if
the condition $\omega_g=\omega_l$ is satisfied), are given by the following equations:
\begin{equation}
H_{\mathrm{eff}}(t)  = \hbar \omega_g\mathbb{I}_{3}-\hbar\Theta^{\dagger}(t)\Theta(t),\label{OmegadaggerOmega}
\end{equation}
with
\begin{eqnarray}
\Theta(t) & \equiv & \left[\begin{array}{ccc}
\frac{\Omega_{1,1}(t)}{\sqrt{\Delta_{1}}} & \frac{\Omega_{1,0}(t)}{\sqrt{\Delta_{1}}} & \frac{\Omega_{1,-1}(t)}{\sqrt{\Delta_{1}}}\\
\frac{\Omega_{2,1}(t)}{\sqrt{\Delta_{2}}} & \frac{\Omega_{2,0}(t)}{\sqrt{\Delta_{2}}} & \frac{\Omega_{2,-1}(t)}{\sqrt{\Delta_{2}}}\\
\frac{\Omega_{3,1}(t)}{\sqrt{\Delta_{3}}} & \frac{\Omega_{3,0}(t)}{\sqrt{\Delta_{3}}} & \frac{\Omega_{3,-1}(t)}{\sqrt{\Delta_{3}}}
\end{array}\right],\label{Theta}
\end{eqnarray}
where it holds $\sqrt{\Delta_{s}}=i\sqrt{-\Delta_{s}},$ since all detunings are negative, and we choose the parameters so that
\begin{eqnarray}
    \Theta(t) & = & \Theta^{\dagger}(t) =\Upsilon(t).\label{relation}
\end{eqnarray}
Thus, for a given gate Hamiltonian as in Eq.~\eqref{Hd}, we calculate
the elements of the operator $\Upsilon(t)$ according to Eq.~\eqref{Upsilon},
and then we must generate fields in the laboratory resulting in Rabi
frequencies such that Eq.~\eqref{relation} is satisfied. It follows that one needs to generate fields such that the diagonal frequencies $\Omega_{1,1}(t)$, $\Omega_{2,0}(t)$, and $\Omega_{3,1}(t)$ are purely imaginary while the off-diagonal frequencies must satisfy
\begin{eqnarray}
    \frac{\Omega_{2,1}(t)}{\sqrt{\Delta_2}} &=&\frac{\Omega_{1,0}^{\ast}(t)}{\left(\sqrt{\Delta_1}\right)^{\ast}},\quad \frac{\Omega_{3,1}(t)}{\sqrt{\Delta_3}} = \frac{\Omega_{1,-1}^{\ast}(t)}{\left(\sqrt{\Delta_1}\right)^{\ast}},\nonumber \\
    \frac{\Omega_{2,0}(t)}{\sqrt{\Delta_3}} &=& \frac{\Omega_{2,-1}^{\ast}(t)}{\left(\sqrt{\Delta_2}\right)^{\ast}}.\label{condition}
\end{eqnarray}
As a consequence, even if the nine complex Rabi frequencies are by definition in general independent \cite{napolitano2019protecting}, the parameters of the external fields are chosen in such a way that three of the off-diagonal frequencies are linked to the other three according to Eq.~\eqref{condition}. Notice that the three detunings involved in the procedure are constant and independent. 

In the next section we study several one-qudit quantum gates and show simulations of the GCDD process.

\section{Results}\label{Sec:Rabi}
Here we explore several applications of the GCDD procedure in the case of a qutrit, by focusing mainly on the required time-dependence for the various Rabi frequencies involved in the decoupling protocol. In all the applications the noise involves both damping and dephasing effects. 
In Sec.~\ref{Subsec:noise} we briefly describe the noise model considered in our numerical simulations.
In Sec.~\ref{Subsec:Hadamard} we calculate the Rabi frequencies required to generate and protect the action of a qutrit generalized Hadamard gate. In particular, the effect of noise is studied both at zero and nonzero temperature. In Sec.~\ref{Subsec:twogates} we consider two additional gates, which can be thought as a generalization of the usual $X$ and $Z$ qubit gates, and simulate their protection using the GCDD method. In Sec.~\ref{Subsec:universal} we verify that it is possible to emulate any quantum gate represented by a generic $\mathrm{SU}(3)$ matrix without the need of a universal set of gates. In Sec.~\ref{Subsec:transitions} we study how the Rabi frequencies behave when we finish applying a quantum gate and start applying another one in sequence. Finally, in Sec.~\ref{Subsec:example} we show an example of how the GCDD can be used to execute a very simple one-qutrit algorithm, described in Ref.~\cite{gedik2015computational}.
 
\subsection{Noise model and fidelity}\label{Subsec:noise}
 
In our model the dissipative effects are  assumed to derive from the interaction
between our qutrit system and two identical  bosonic thermal baths, respectively responsible for amplitude damping and dephasing noise. A complete description of 
how to treat the dissipative dynamics can be found in the main text of Ref.~\cite{napolitano2019protecting} and in its Appendix C.  The bosonic fields are characterized by the same spectral density  $J\left(\omega \right)=\alpha ^{2}\omega \exp \left(-\omega/\omega_{c}\right)$, where $\alpha $ is a dimensionless constant linked to the noise strength and $\omega_{c}$ is the cut-off frequency which is also connected to the correlation time of the boson fields $\tau_{c}$, as $\tau_{c}\equiv 2\pi /\omega_{c}$. We stress that amplitude damping and dephasing noises are widely considered in studies concerning open quantum systems and here are employed  as a theoretical model, not connected to a specific experimental platform, to test our control procedure. 
 
Using the state nomenclature of Fig.~\ref{Fig1}, the amplitude damping noise is due to emission and absorption processes corresponding to atomic transitions between the states $\left| \pm 1\right\rangle$ and $\left| 0\right\rangle$, while the dephasing results in destroying the phases between $\left| \pm 1\right\rangle$ and $\left| 0\right\rangle$. In fact, in Ref.~\cite{napolitano2019protecting} the dephasing noise does not couple the states $\ket{1}$ and $\ket{-1}$ directly. In this work, we have added a direct coupling between these states in the dephasing noise, that is, we have included a term proportional to $\ketbra{-1}{-1} - \ketbra{1}{1}$ in the interaction Hamiltonian with the dephasing bath (see Appendix C of Ref.~\cite{napolitano2019protecting}).
 
Differently from Ref.~\cite{napolitano2019protecting}, in our simulations we assume for the cut-off frequency, $\omega_c/( 2 \pi)= 10^6$ Hz. This choice implies for the bath correlation time, $\tau_c = 1$ $\mu$s. Since we don't consider a specific setup as source of the noise, the value chosen for $\omega_c$ is just a reasonable one leading to a correlation time much smaller than the coherence times involved in typical experiments with atomic systems, which are the order of a second or longer \cite{PhysRevA.94.013427, PhysRevLett.92.203005, PhysRevLett.105.020401}. Besides, we take all the coupling constants appearing in the interaction terms with the environment all equal between them, here including the new coupling constant appearing in the added term for the dephasing noise. We name $\lambda$ this common coupling constant. In order to clarify the notation adopted here with respect to Ref.~\cite{napolitano2019protecting}  we remark the following. In Ref.~\cite{napolitano2019protecting} we named as effective coupling parameter the quantity $\bar{\lambda}=\lambda \alpha$. However, in the master equation the final quantity giving the effective strength of the noises is $\lambda_{\mathrm{me}}=\bar{\lambda}\alpha=\lambda \alpha^2$. In this sense, in the figures of Ref.~\cite{napolitano2019protecting} giving the value of $\bar{\lambda}$ was not sufficient to obtain the value of $\lambda_{\mathrm{me}}$ effectively used in the master equation. In order to do that, it is enough to consider that it was like if we assumed $\lambda= 1$. In this way the value $\bar{\lambda}=0.1$ used in Ref.~\cite{napolitano2019protecting} corresponds here to $\lambda_{\mathrm{me}}=0.01$.

The impact of our protection scheme is here quantified by means of the fidelity measuring how close are the states obtained, respectively, by the dissipative dynamics associated to the total Hamiltonian of Eq.~\eqref{eq:Htot} (treated by means of a Redfield master equation \cite{napolitano2019protecting}) and by the noise-free dynamics governed by the Hamiltonian $H_{\mathrm{lab}}(t)$ of Eq.~\eqref{Hlab}. The latter dynamics gives after a time $\tau$ the	same output state of the ideal noise-free dynamics governed by the gate Hamiltonian $H_G$ of Eq.~\eqref{Hd} (i.e., the dynamics in the absence of the  protective scheme). In our study we also consider the dissipative dynamics in the absence of the protective scheme, which is obtained by turning off the control Hamiltonian $H_c(t)$. Since the two baths are always assumed to be identical we refer to them using the term bath as singular.

Concerning the fidelity, it is defined for two arbitrary states $\rho$ and $\sigma$ as $[\mathrm{Tr}\{\sqrt{\sqrt{\rho} \sigma \sqrt{\rho}}\}]^2$. In order to properly characterize the performance of an arbitrary gate $U$, we consider the average over a suitable set of initial states $\rho(0)$. According to Ref.~\cite{nielsen2002simple}, we can define the average one-qudit gate fidelity as
\begin{eqnarray}\label{avgfid-d}
    \overline{F}(U,\mathcal{E}) & = & \frac{1}{d+1} \nonumber \\
    &&+ \frac{1}{d^2\left(d+1\right)} \sum_j \Tr\left\{ U U_j^\dagger U^\dagger \mathcal{E}\left(U_j^\dagger\right)\right\},
\end{eqnarray}
where the $U_j$ form a unitary operator basis for the qudit and $\mathcal{E}$ is the map associated to the master equation. The $U_j$ must be orthogonal with respect to the Hilbert-Schmidt inner product, that is, they are such that $\Tr\{U_j^\dagger U_k\} = d \delta_{jk}$, where $\delta_{jk}$ is the Kronecker delta. In our qutrit case, a natural choice is represented by the Gell-Mann matrices $\lambda_j$, and so we set $U_0$ as the qutrit identity matrix $\mathbb{I}_{3}$, and $U_j = \sqrt{3/2} \lambda_j$. The average fidelity for a one-qutrit gate can then be calculated as
\begin{eqnarray}\label{avgfid-3}
    \overline{F}(U,\mathcal{E}) = \frac{1}{3} + \frac{1}{24} \sum_{j=1}^8 \Tr\left\{ U \lambda_j U^\dagger \mathcal{E}\left(\lambda_j\right) \right\}.
\end{eqnarray}
Notice that although individually the Gell-Mann matrices do not represent physical density operators, the above considered basis can be used to represent any physical state. It follows that the master equation can be used to compute $\mathcal{E}(\lambda_j)$ for any $j$.

\subsection{The Hadamard gate}\label{Subsec:Hadamard}
Let us consider the single-qudit gate known as the Hadamard gate for $d=3$, which we have also studied in Ref.~\cite{napolitano2019protecting}. This gate maps a state of the computational basis into a superposition
of all states and its unitary matrix representation can be written as (here and in the following the matrices are given in the ground-state subspace basis $\{\ket{-1}$, $\ket{0}$, $\ket{1}\}$)
\begin{eqnarray}\label{uhadamard}
U_H & = & -\frac{i}{\sqrt{3}} \left[\begin{array}{ccc}
1 & 1 & 1 \\
1 & e^{i\frac{2\pi}{3}} & e^{-i\frac{2\pi}{3}} \\
1 & e^{-i\frac{2\pi}{3}} & e^{i\frac{2\pi}{3}}
\end{array}\right].\label{UH}
\end{eqnarray}
This unitary satisfies
\begin{equation}
    U_H = \exp(-i \frac{H}{\hbar} \tau),
\end{equation}
where $H$ is the Hamiltonian associated to the Hadamard gate ($U_H$ gate). We can invert this relation by using the matrix logarithm in both sides of the equality to obtain $H$. Then, taking the gate Hamiltonian of Eq.~\eqref{Hd}, and combining it with the definition of
the $H_{L}^{\prime}$ and $H_{F}^{\prime}$ operators, we get the matrix which corresponds to the square of $\Upsilon(t)$, using Eq.~\eqref{Upsilon}.
Unfortunately the resulting matrix is, in general, not analytically diagonalizable for a generic time instant $t$. The numerical diagonalization however is straightforward. Therefore in order to calculate $\Upsilon$ for a specific time $t_i$, we first need to evaluate $\Upsilon^2(t_i)$ and then compute $\Upsilon(t_i)$. 
\begin{figure}
\begin{centering}
\includegraphics[width=0.48\textwidth]{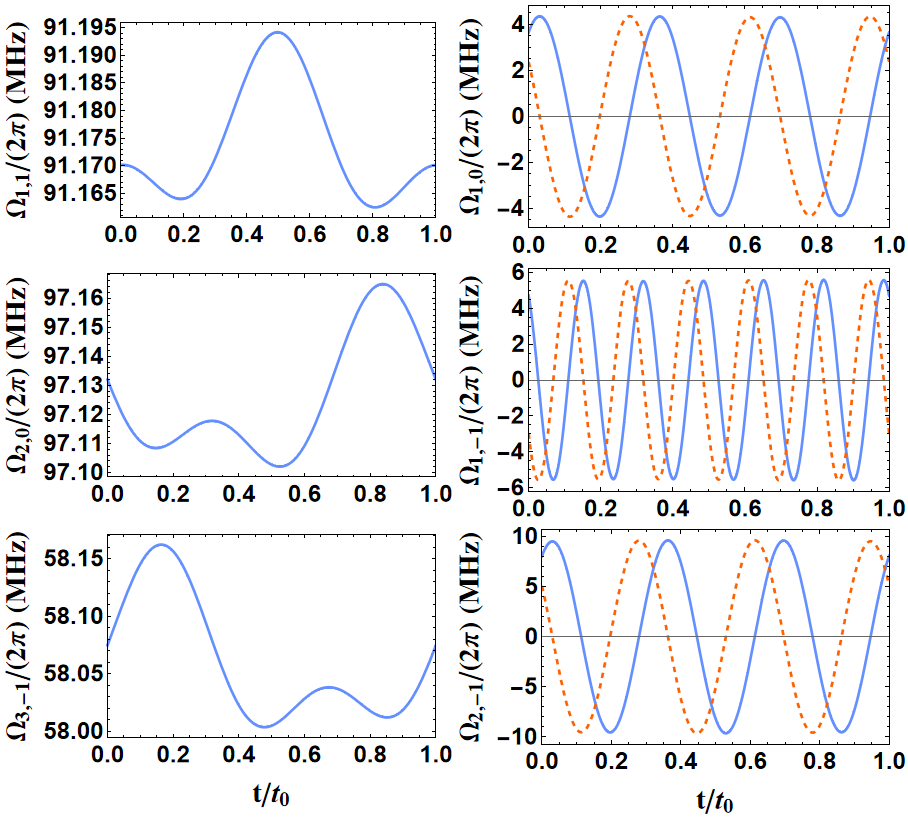}
\end{centering}
\caption{\label{Hadamard-Rabis} Numerical solutions for six of the nine Rabi frequencies [the three remaining Rabi frequencies are directly linked to the ones represented on the right column according to Eq.~\eqref{condition}], as functions of time, used to protect the action of the Hadamard gate $U_H$ on a single qutrit. The following parameters and the styles of the curves are the same as in Figs.~\ref{X-Rabis}, \ref{Z-Rabis}, \ref{Ua-Rabis}, \ref{Ub-Rabis}, \ref{Uc-Rabis}, \ref{f2-Rabis}, and \ref{f5-Rabis}. The chosen detunings are: $\Delta_1/(2\pi) = -1$ GHz, $\Delta_2/(2\pi) = -2$ GHz, and $\Delta_3/(2\pi) = -3$ GHz. The imaginary parts of the Rabi frequencies are shown with solid lines, while the real parts with dashed lines. Due to the fact that $U_c(t)$ is periodic, the plots consist of only one period $t_0$, in units of the gate time $\tau$. The control period is chosen to be $t_0 = \tau/75$, and the value for the gate time is $\tau = 2\pi \times 10^{-5}$ s. Notice that with this choice of the gate time, the maximum value of $|\Omega_{r,s}|/(2\pi)$ is of the order of $100$ MHz.}
\end{figure}

The numerical solutions for the Rabi frequencies necessary to implement the Hadamard gate are shown in Fig.~\ref{Hadamard-Rabis}. In particular, here and in other figures, we consider the Rabi frequencies instead of the matrix elements of the operator $\Upsilon^2(t)$, in order to focus our analysis on the values required for these control parameters in our procedure. Of course the corresponding values for $\Upsilon^2(t)$ and then for $H_{\mathrm{lab}}(t)$ can be easily deduced. As one can see, the shape of the numerical solutions resemble to smooth trigonometric functions. We stress that Rabi frequencies $|\Omega_{r,s}|/(2\pi)$ up to values of the order of  $100$ MHz are currently experimentally achievable~\cite{whiting2016direct}. One can see from Fig.~\ref{Hadamard-Rabis} that by choosing a gate time $\tau = 2 \pi \times 10^{-5}$ s the maximum value for the Rabi frequencies required in our control protocol is of the order of $100$ MHz.

It is also relevant to mention that these shapes do not change for different values of $t_0$, when the time is expressed in units of the control period. The only difference would be the magnitude of the Rabi frequencies. From Eq.~\eqref{Upsilon}, it is possible to conclude that the shape only depends on the gate being implemented, since the control fields $H_L$ and $H_F$ are the same in all cases.
This leads us to conclude that it should be possible to effectively implement the required time dependency of the Rabi frequencies in a laboratory 
through the use of electro-optic modulators, for phase control, combined with Mach-Zehnder interferometers. To be more specific, the control over the electric field phase can be realized by controlling the difference of electric potential applied inside the electro-optical modulator. This creates a time dependency on the refraction index, which in turn, will create a time dependency on the electric field phase. If the laser beam is split in two and only one of the beams goes through the modulator, we create a pair of beams with a difference in its relative phase that can be completely controlled. Finally, by combining both beams again, we can create interference, meaning we can translate the time dependency of the relative phase into control of the amplitude of the recombined beams~\cite{govindagrawal2010}. Since the time-dependent Rabi frequencies are proportional to the laser amplitude, then it should be possible to generate them through the process described.

\begin{figure}
	\begin{centering}
		\includegraphics[width=0.48\textwidth]{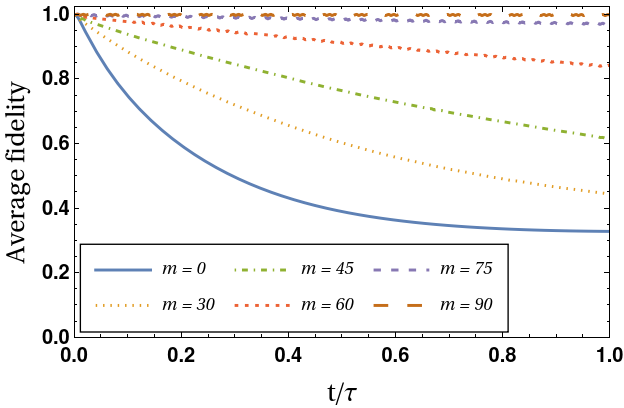}
	\end{centering}
	\caption{\label{Hadamard-ruido} Numerical solutions to estimate how much  the GCDD procedure protects the amplitude damping and dephasing noises during the action of the Hadamard gate $U_H$ (the time $t$ is given  in units of the gate time $\tau = 2 \pi \times 10^{-5}$ s), for different control periods $t_0$. The quantity $m$ in the legend is defined as $m \equiv \tau/t_0$. The reference case in which there is no protection is also shown ($m=0$). The temperature is $T = 300$ K, the coupling constants between the system and the baths are all equal, and the effective strength of the noises is $\lambda_{\text{me}} = 10^{-10}$. The chosen cut-off frequency, $\omega_c /( 2 \pi) = 10^6$ Hz, is equal to the one used in Figs.~\ref{Hadamard-ruido-zero}, \ref{sequential_noise}, and \ref{superposed}. The bath correlation time $\tau_c$ is equal to $\tau_c = 2\pi/\omega_c=1$ $\mu$s.}
\end{figure}

To show that with the use of these Rabi frequencies one can protect the qutrit from noise
while also applying the desired Hadamard gate, we have performed simulations of different kinds of dynamics for several values of the control period $t_0$. The average fidelity as a function of time is shown in Fig.~\ref{Hadamard-ruido} in the case of a bath at room temperature, using  $T= 300$ K. Differently from Ref.~\cite{napolitano2019protecting}, as in Fig.~\ref{Hadamard-Rabis} and in the subsequent figures, $t_0$ is given in units of the gate time $\tau$ (we use the quantity $m \equiv \tau/t_0$), instead of the bath correlation time $\tau_c$, because the protection scheme does not depend on the characteristics of the noise. This choice is also due to the fact that from an experimental point of view, the reference time one has access to is the gate time. 
One can notice that the average gate fidelity becomes higher with respect to the case without protection the shorter the control period used is. Therefore, the solutions giving the Rabi frequencies presented in Fig.~\ref{Hadamard-Rabis} can indeed provide protection from noise while also executing the Hadamard gate to our qutrit. We observe that it is easier to protect the qudit the higher $\tau_c$ is.

\begin{figure}
	\begin{centering}
		\includegraphics[width=0.48\textwidth]{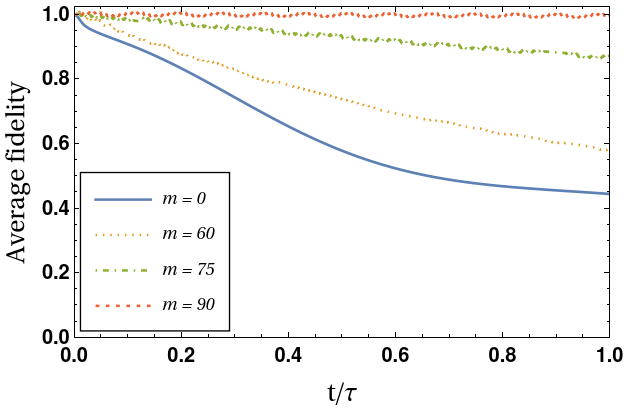}
	\end{centering}
	\caption{\label{Hadamard-ruido-zero} Impact of the GCDD procedure for different values of $t_0$ (we use the quantity $m = \tau/t_0$), for the Hadamard gate $U_H$ at zero temperature (with both amplitude damping and dephasing noises). The coupling constant used is $\lambda_{\text{me}} = 5 \times 10^{-3}$ and, as in Fig.~\ref{Hadamard-ruido}, we use $\tau = 2 \pi \times 10^{-5}$ s, and $\omega_c /( 2 \pi)= 10^6$ Hz.}
\end{figure}

We have also considered the case in which the bath can be assumed to be at zero temperature. In this limit we have only vacuum-noise effects in terms of dephasing and amplitude damping. The average fidelity as a function of time, for different control periods, in the case of vacuum noise is shown in Fig.~\ref{Hadamard-ruido-zero}. Notice that, even though the fidelity in the absence of protection drops only to around $0.45$ instead of $1/3$, we still need $m$ of the same order of those used in the case where there is also thermal noise (see Fig.~\ref{Hadamard-ruido}) in order to achieve fidelities above $0.99$. 

In both Figs.~\ref{Hadamard-ruido} and \ref{Hadamard-ruido-zero} the cut-off frequency is $\omega_c /(2 \pi)= 10^6$ Hz, which implies a bath correlation time of $1$ $\mu$s. We remark that bath correlation times of this order of magnitude or even longer have been experimentally observed through noise spectral-density analysis~\cite{alvarez2011measuring, szankowski2017environmental,  zhang2020discrimination}. Since the control period must be shorter than the correlation time, the protection schemes working for $\tau_c = 1$ $\mu$s can also be exploited for higher values of the bath correlation time. For the thermal case the effective coupling parameter is $\lambda_{\text{me}} = 10^{-10}$ while for the vacuum case it holds $\lambda_{\text{me}} = 5 \times 10^{-3}$. These parameters have been chosen so that the protection scheme would be effective already for $\tau/t_0 \le 100$. For stronger values of coupling and/or smaller correlation times, the method would require faster control fields, meaning $\tau/t_0 > 100$. As a consequence, we would need Rabi frequencies such that $|\Omega_{r,s}|/(2\pi)$ is higher than $100$ MHz, and this would be very challenging from an experimental point of view. We notice that the same Rabi frequencies that arise from the GCDD method are used to protect the qutrit against both vacuum and thermal noise. This aspect has been already treated in the final comment of Sec.~\ref{Sec:GCDD}.

\subsection{Two additional gates}\label{Subsec:twogates}
We have performed the same procedure for two other one-qutrit gates, which we name $U_X$ and $U_Z$. Their unitary matrix representations are written as follows:
\begin{eqnarray}\label{ux}
U_{X} & = & \left[\begin{array}{ccc}
0 & 0 & 1 \\
1 & 0 & 0 \\
0 & 1 & 0
\end{array}\right],\label{A0}
\end{eqnarray}
\begin{eqnarray}\label{uz}
U_{Z} & = & \left[\begin{array}{ccc}
1 & 0 & 0 \\
0 & e^{i\frac{2\pi}{3}} & 0 \\
0 & 0 & e^{-i\frac{2\pi}{3}}
\end{array}\right].\label{P0}
\end{eqnarray}
These two gates can be thought as extensions of the usual $X$ and $Z$ one-qubit gates to the three-dimensional case. $U_X$ flips the quantum states cyclically while $U_Z$ applies phases of $\pm 2\pi/3$ to the states $\ket{0}$ and $\ket{1}$, respectively, and does nothing to the state $\ket{-1}$. 

\begin{figure}
	\begin{centering}
		\includegraphics[width=0.48\textwidth]{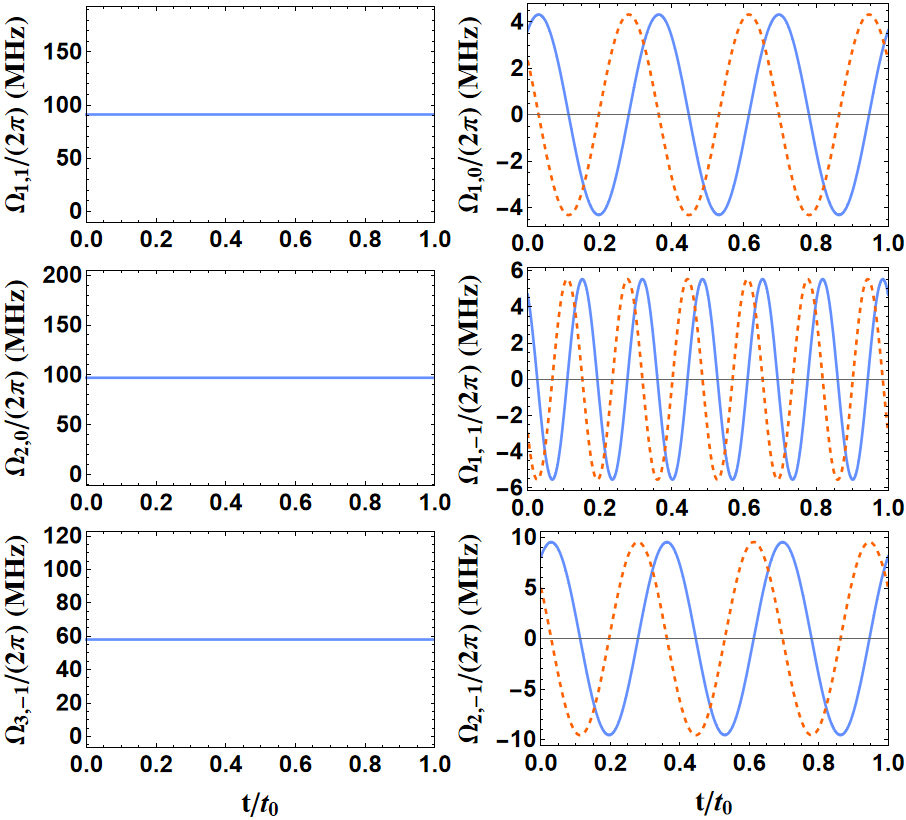}
	\end{centering}
	\caption{\label{X-Rabis} Same plots and parameters as in Fig.~\ref{Hadamard-Rabis}, but for the $U_X$ gate.}
\end{figure}

\begin{figure}[h!]
	\begin{centering}
		\includegraphics[width=0.48\textwidth]{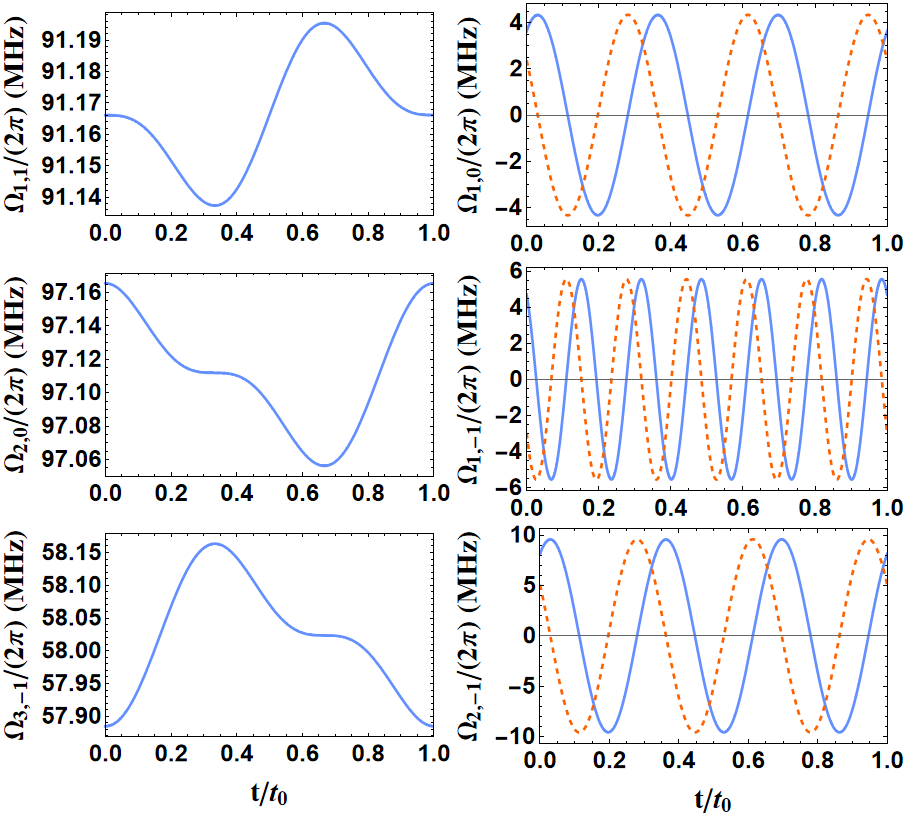}
	\end{centering}
	\caption{\label{Z-Rabis} Same plots and parameters as in Fig.~\ref{Hadamard-Rabis}, but for the $U_Z$ gate.}
\end{figure}

Similarly to the case of the Hadamard gate, we present the Rabi frequencies for the $U_X$ and $U_Z$ gates in Figs.~\ref{X-Rabis} and \ref{Z-Rabis}, respectively.
We are always assuming that the control period $t_0$ is chosen to be sufficiently shorter than the bath correlation time $\tau_c$. The GCDD simulations of the above three distinct one-qutrit gates (not shown for the $U_X$ and $U_Z$ gates) strongly suggest that this method might be used to generate any one-qutrit gate while also protecting it from external noise. In particular, as far as the gate dynamics studied here indicate, by choosing $t_{0} = \tau/90$, we can achieve fidelity values above $0.99$. 

\subsection{Beyond an universal set of quantum gates}\label{Subsec:universal}
It is a well-known result that a finite set of one-qudit gates combined with an entangling gate can be used to approximate any  multi-qudit quantum gate within arbitrary precision \cite{dawson2005solovay}. This is a useful concept since there are quantum gates easier to implement than others. In our present atomic qutrit model, we have shown that in order to apply a protected quantum gate all is needed is to generate specific time-dependent Rabi frequencies. Therefore, depending on the precision the Rabi frequencies can be controlled in laboratory, it may be possible to generate arbitrarily complex gates without the need of decomposing them into simpler gates.

To illustrate this possibility, we consider three different quantum gates that are not particularly useful in any direct algorithm, but instead, illustrate what a general single-qutrit gate looks like. It is possible to write any $3\times 3$ unitary matrix using a set of eight parameters given by $p = \{\theta_1, \theta_2, \theta_3, \phi_1, \phi_2, \phi_3, \phi_4, \phi_5\}$, that is
\begin{eqnarray}\label{general_U}
    U = \begin{bmatrix}
    u_{00}(p) & u_{01}(p) & u_{02}(p) \\ u_{10}(p) & u_{11}(p) & u_{12}(p) \\ u_{20}(p) & u_{21}(p) & u_{22}(p)
    \end{bmatrix}.
\end{eqnarray}
The parameters are such that $\theta_j \in [0,\pi]$ and $\phi_j \in [0,2\pi[$. The explicit form of the entries $u_{jk}(p)$ are given by \cite{bronzan1988parametrization}
\begin{eqnarray}
\begin{split}
    u_{00}(p) &= \cos{\theta_1} \cos{\theta_2} e^{i \phi_1}, \\
    u_{01}(p) &= \sin{\theta_1} e^{i \phi_3}, \\
    u_{02}(p) &= \cos{\theta_1} \sin{\theta_2} e^{i \phi_4}, \\
    u_{10}(p) &= \sin{\theta_2} \sin{\theta_3} e^{-i\phi_4 -i\phi_5} \\ &\quad\, - \sin{\theta_1} \cos{\theta_2} \cos{\theta_3} e^{i\phi_1 + i\phi_2 - i\phi_3}, \\
    u_{11}(p) &= \cos{\theta_1} \cos{\theta_3} e^{i\phi_2}, \\
    u_{12}(p) &= - \cos{\theta_2} \sin{\theta_3} e^{-i\phi_1 -i\phi_5}, \\ &\quad\, - \sin{\theta_1} \sin{\theta_2} \cos{\theta_3} e^{i\phi_2 - i\phi_3 + i\phi_4}, \\
    u_{20}(p) &= - \sin{\theta_1} \cos{\theta_2} \sin{\theta_3} e^{i\phi_1 -i\phi_3+i\phi_5} \\ &\quad\, - \sin{\theta_2} \cos{\theta_3} e^{-i\phi_2 - i\phi_4}, \\
    u_{21}(p) &= \cos{\theta_1} \sin{\theta_3} e^{i\phi_5}, \\
    u_{22}(p) &= \cos{\theta_2} \cos{\theta_3} e^{-i\phi_1 -i\phi_2} \\ &\quad\, - \sin{\theta_1} \sin{\theta_2} \sin{\theta_3} e^{-i\phi_3 + i\phi_4 + i\phi_5}.
\end{split}
\end{eqnarray}
This parametrization might seem confusing and dense, but, in the context of numerical simulation, it is simpler than the usual parametrization using the Lie algebra generators~\cite{hall2013lie}.

\begin{figure}
	\begin{centering}
		\includegraphics[width=0.48\textwidth]{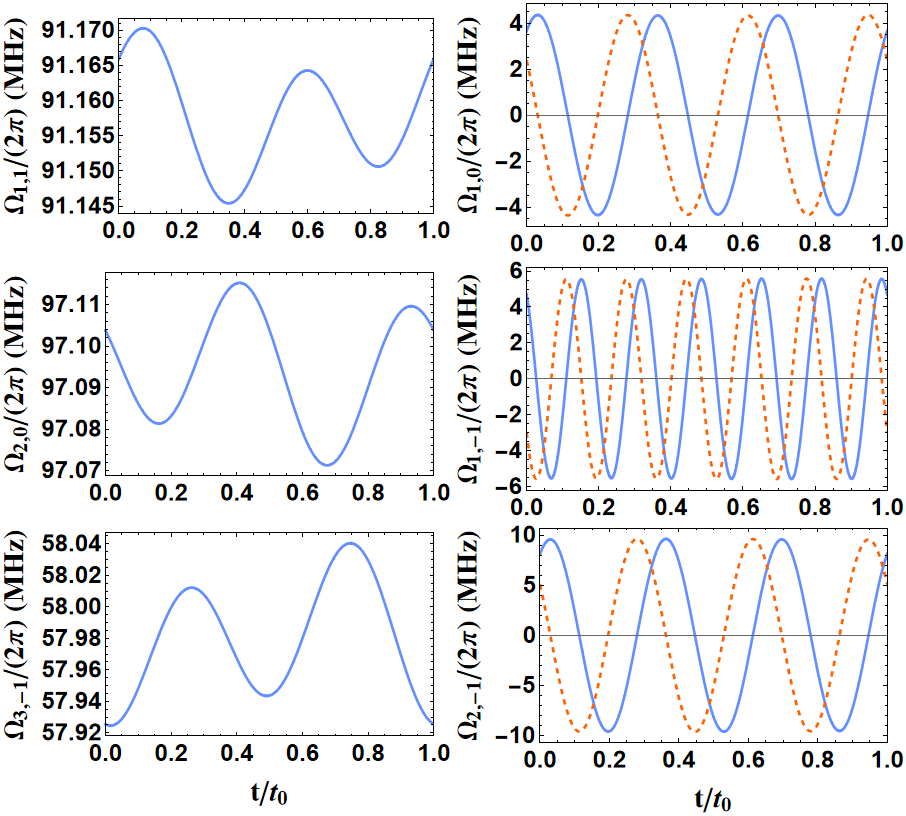}
	\end{centering}
	\caption{\label{Ua-Rabis}{Same plots and parameters as in Fig.~\ref{Hadamard-Rabis}, but for the $U_{\alpha}$ gate.}}
\end{figure}

\begin{figure}
	\begin{centering}
		\includegraphics[width=0.48\textwidth]{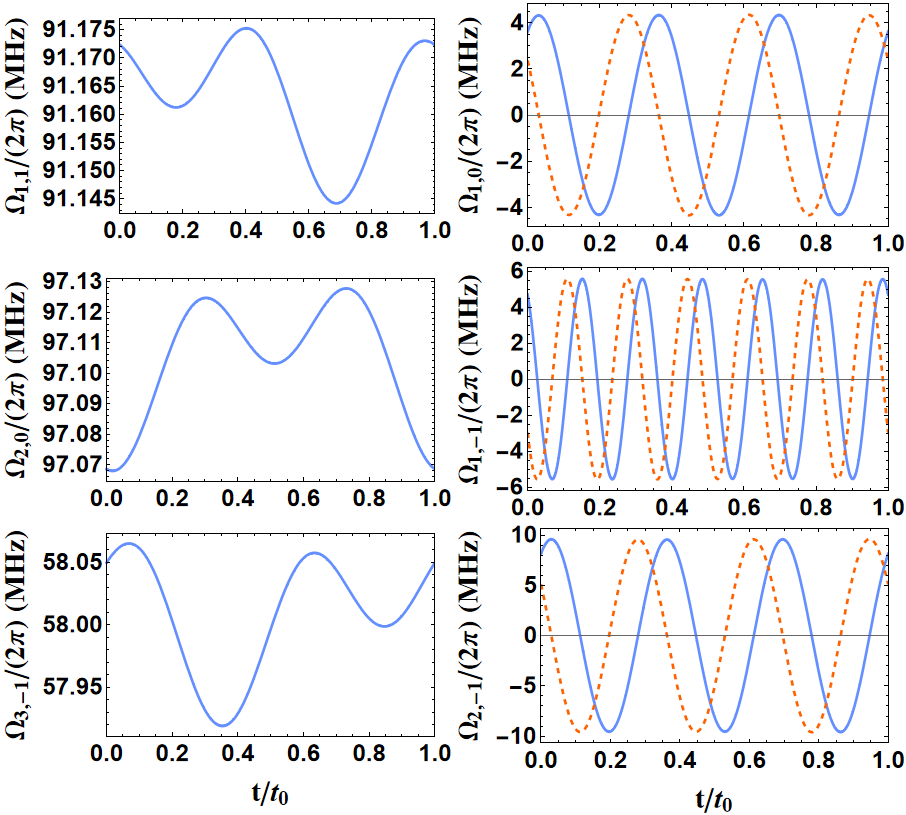}
	\end{centering}
	\caption{\label{Ub-Rabis}{Same plots and parameters as in Fig.~\ref{Hadamard-Rabis}, but for the $U_{\beta}$ gate.}}
\end{figure}

\begin{figure}
	\begin{centering}
		\includegraphics[width=0.48\textwidth]{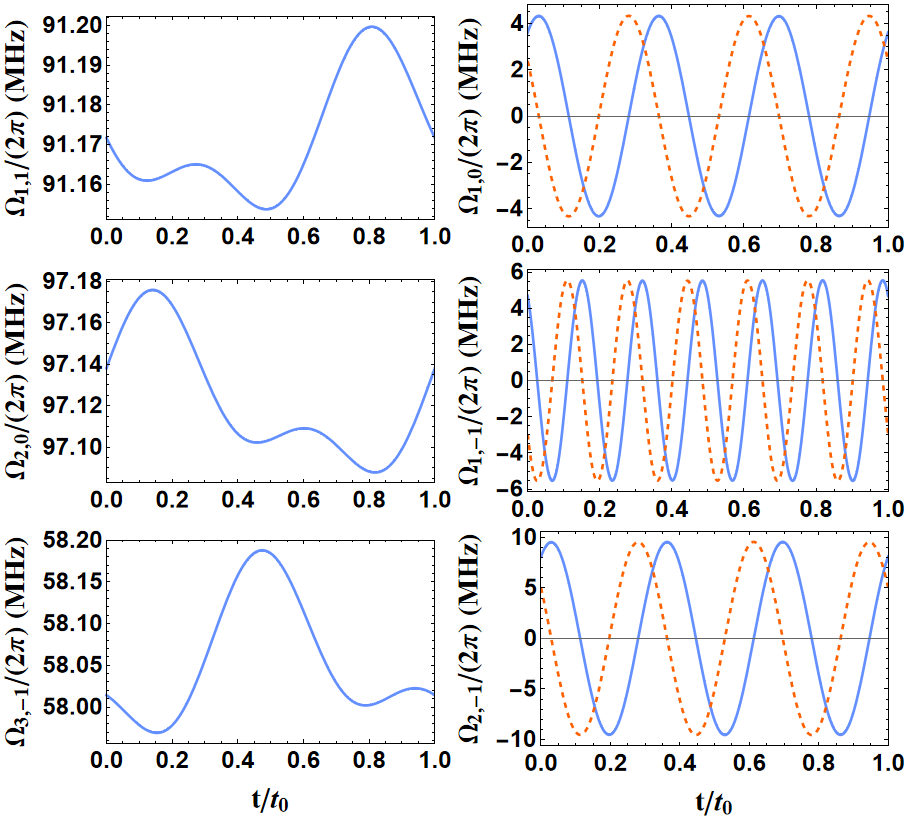}
	\end{centering}
	\caption{\label{Uc-Rabis}{Same plots and parameters as in Fig.~\ref{Hadamard-Rabis}, but for the $U_{\gamma}$ gate.}}
\end{figure}

We have generated pseudo-random values for the eight parameters to build up three different one-qutrit gates. The three sets generated are:
\begin{eqnarray}\label{Set_a}
\begin{split}
    \theta_1^{(\alpha)} = 0.924939 \times \pi , \;\;\; \phi_1^{(\alpha)} &= 0.773065 \times \pi , \\
    \theta_2^{(\alpha)} = 0.109418 \times \pi , \;\;\; \phi_2^{(\alpha)} &= 1.623579 \times \pi , \\
    \theta_3^{(\alpha)} = 0.424271 \times \pi , \;\;\; \phi_3^{(\alpha)} &= 0.562634 \times \pi , \\
    \phi_4^{(\alpha)} &= 0.966021 \times \pi, \\
    \phi_5^{(\alpha)} &= 0.632344 \times \pi,
\end{split}
\end{eqnarray}
\begin{eqnarray}\label{Set_b}
\begin{split}
    \theta_1^{(\beta)} = 0.363719 \times \pi , \;\;\; \phi_1^{(\beta)} &= 0.254886 \times \pi , \\
    \theta_2^{(\beta)} = 0.098480 \times \pi , \;\;\; \phi_2^{(\beta)} &= 1.419032 \times \pi , \\
    \theta_3^{(\beta)} = 0.420599 \times \pi , \;\;\; \phi_3^{(\beta)} &= 1.336089 \times \pi , \\
    \phi_4^{(\beta)} &= 1.284848 \times \pi, \\
    \phi_5^{(\beta)} &= 0.837395 \times \pi,
\end{split}
\end{eqnarray}
\begin{eqnarray}\label{Set_C}
\begin{split}
    \theta_1^{(\gamma)} = 0.539910 \times \pi , \;\;\; \phi_1^{(\gamma)} &= 1.491107 \times \pi , \\
    \theta_2^{(\gamma)} = 0.947045 \times \pi , \;\;\; \phi_2^{(\gamma)} &= 0.950144 \times \pi , \\
    \theta_3^{(\gamma)} = 0.179982 \times \pi , \;\;\; \phi_3^{(\gamma)} &= 1.780184 \times \pi , \\
    \phi_4^{(\gamma)} &= 0.919544 \times \pi, \\
    \phi_5^{(\gamma)} &= 1.573260 \times \pi.
\end{split}
\end{eqnarray}
Plugging these values into Eq.~\eqref{general_U} yields three one-qutrit gates which we name $U_{\alpha}$, $U_{\beta}$, and $U_{\gamma}$, respectively. Using the same procedure described for the gates $U_H$, $U_X$, and $U_Z$, we calculate the time dependence of the Rabi frequencies as the ones shown in Figs.~\ref{Ua-Rabis}-\ref{Uc-Rabis}. As for the gates $U_H$, $U_X$, and $U_Z$, also these three general gates require a time dependency for the Rabi frequencies that resemble simple trigonometric functions. This suggests that, depending on the precision at which the Rabi frequencies can be controlled in the laboratory, it might be possible to implement general one-qutrit gates without the need of decomposing them into simpler, universal gates. 


Of course, this is just a speculative conclusion over the numerical results. However, if in a practical implementation attempt of these gates one could verify that it is not necessary to decompose them into simpler gates, this could mean a reasonable improvement in the complexity in algorithms involving complicated one-qutrit (or qudit) gates.

\subsection{Rabi frequencies in gate transitions}\label{Subsec:transitions}

It is pertinent to ask how the Rabi frequencies change when different gates are applied in sequence from one to the next (let us say from $G_1$ to $G_2$). The only quantity that must change in the $\Upsilon(t)$ matrix is the operator $G$. We must then only make the substitution
\begin{eqnarray}
    G & \rightarrow & G_1 \theta(\tau-t) + G_2 \theta(t-\tau),
\end{eqnarray}
where $\theta$ is the Heaviside step function. Obviously this would mean that the Rabi  frequencies would need to be instantaneously discontinuous at the transition of the gates, which is not implementable in a laboratory. To overcome this problem we can replace the Heaviside step function with a smooth function defined as
\begin{eqnarray}
    \theta_s(t) & = & \frac{1}{\exp(-\mu t)+1},
\end{eqnarray}
that becomes the Heaviside function in the limit $\mu \rightarrow \infty$. 

\begin{figure}
	\begin{centering}
		\includegraphics[width=0.48\textwidth]{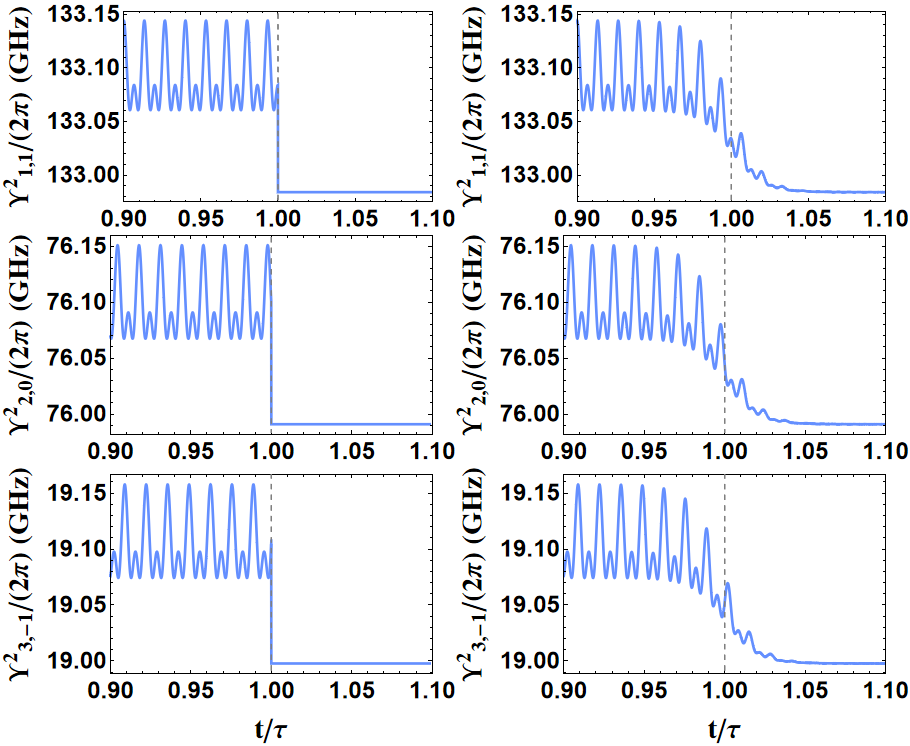}
	\end{centering}
	\caption{\label{rabi_transition} Diagonal elements of the operator $\Upsilon^2(t)$, as functions of time, with a transition between the application of the Hadamard gate $U_H$ and the period where the memory is kept (corresponding to the application of the identity gate $\mathbb{I}_3$), using the Heaviside $\theta$ function (left column) and the smooth function $\theta_s$ (right column) with $\mu=100/\tau$. Same parameters as in Fig.~\ref{Hadamard-Rabis}.}
\end{figure}

\begin{figure}[t!]
	\begin{centering}
		\includegraphics[width=0.48\textwidth]{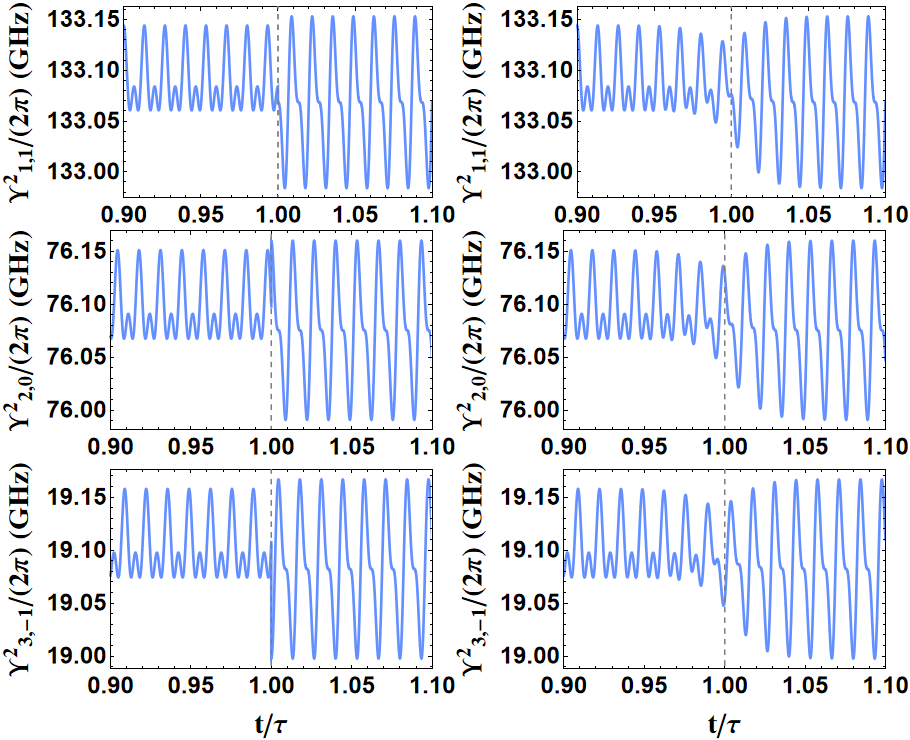}
	\end{centering}
	\caption{\label{HZ_transition} Same plots and parameters as in Fig.~\ref{rabi_transition}, with a transition between the Hadamard gate $U_H$ and the $U_Z$ gate.}
\end{figure}

As a first example, we consider the case of a transition between the Hadamard gate and a period where the qutrit state is maintained at its state (this is done by turning on the GCDD protection associated to the case $ H_G = 0$). The time dependence of the diagonal elements of the operator $\Upsilon^{2}(t)$ are shown in Fig.~\ref{rabi_transition} using the functions $\theta$ and $\theta_s$. We stress that differently from previous figures, we preferred to report here the values of some elements of $\Upsilon^{2}(t)$ instead of the corresponding Rabi frequencies, whose values can be easily obtained. By setting the parameter $\mu=100/\tau$ the loss in the final fidelity is of the order $10^{-6}$, a negligible value compared to the natural losses due to the external noise. As a second example, we also consider the transition between the Hadamard and $U_Z$ gates in Fig.~\ref{HZ_transition}. Notice how, in both examples, with just the addition of the $\theta_s$ function, it is possible to make the Rabi frequencies and their first derivative continuous at the time the transition between two quantum gates happens.

\begin{figure}[t!]
	\begin{centering}
		\includegraphics[width=0.48\textwidth]{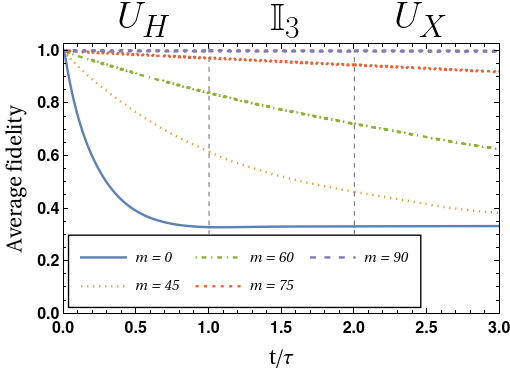}
	\end{centering}
	\caption{\label{sequential_noise} Impact of the GCDD procedure on the average fidelity for different values of $t_0$ (we use the quantity $m = \tau/t_0$), with two gates ($U_H$ and $U_X$) being applied in sequence, separated by an interval where the memory of the state is supposed to be kept by applying the identity gate $\mathbb{I}_3$. As in Fig.~\ref{Hadamard-ruido}, we use  $\tau = 2 \pi \times 10^{-5}$ s, $T = 300$ K,  $\lambda_{\mathrm{me}}= 10^{-10}$, and $\omega_c/(2 \pi)= 10^6$ Hz.}
\end{figure}

We have also simulated the noisy evolution of three operations in sequence: Hadamard gate, followed by a period where the memory is kept (identity operator), followed by the application of the $U_X$ gate given by Eq.~\eqref{ux}. The results are shown in Fig.~\ref{sequential_noise}. By choosing $m=90$ we can achieve a fidelity of around $0.997$ at the end of the process. 

\subsection{Effects of imprecision in the Rabi frequencies}
Here we consider the possibility of imprecision during a practical implementation of the Rabi frequencies. From the figures shown for the several Rabi frequencies one can notice that, specially for the frequencies in the diagonal of the operator $\Upsilon(t)$, the values oscillate with a very small amplitude. The quantity $\Omega_{1,1}(t)$, for example, usually varies within a range of roughly $0.05$ MHz, if we consider a gate time $\tau=2 \pi \times 10^{-5}$ s. For quantum gates such as the $U_X$ gate it is even more critical, as it should be kept constant. So it is relevant to ask what problems would arise in the case where the control over the Rabi frequencies could not be implemented with such precision.

To test this scenario we consider a very simple random function that introduces fluctuations during the numerical calculation of the Rabi frequencies. We may call this function $\rho(t)$, and define it with an interpolation up to third polynomial degree of random points in the range $[-1, 1]$, equally spaced between time intervals $t_\rho$. We then replace the operator $\Upsilon(t)$ with
\begin{eqnarray}
    \Upsilon(t) \rightarrow \Upsilon(t) \left[ 1 + \delta \rho(t) \right],
\end{eqnarray}
where $\delta$ is a real number for modulating the range $[-1, 1]$. We have considered $t_\rho =$ $10^{-2} \times \tau/75$ and $\delta = 0.5 \times 10^{-3}$. 

\begin{figure}
    \centering
    \includegraphics[width=0.48\textwidth]{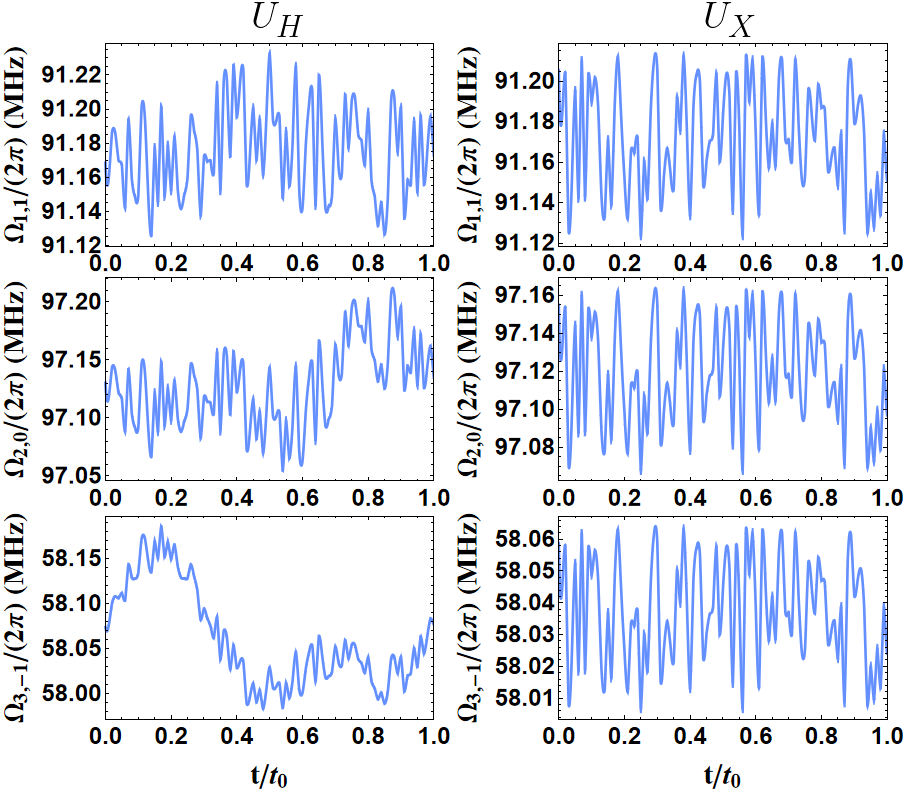}
    \caption{Numerical solutions for the imaginary parts of the diagonal Rabi frequencies (they are purely imaginary) for the Hadamard gate $U_H$ (left column) and the $U_X$ gate (right column), in the case where generation of the control fields are considered to be imprecise. The parameters used are $t_\rho = 10^{-2} t_0$ and $\delta = 0.5 \times 10^{-3}$, with all the other values equal to the ones used in Fig.~\ref{Hadamard-Rabis}. 
    The loss in gate fidelity due to these random fluctuations in the Rabi frequencies are of the order $10^{-5}$, which are roughly in the same range as the loss in fidelity due to gate transitions.}
    \label{imprecision}
\end{figure}

As one can see in Fig.~\ref{imprecision}, where we only show the Rabi frequencies appearing in the diagonal of the operator $\Upsilon(t)$, by choosing these parameters we have these frequencies randomly varying, roughly, in a range between $0.01$ and $0.1$ MHz (we still consider $\tau=2 \pi \times 10^{-5}$ s), and this corresponds to the level of control typically involved in experiments with Rabi frequencies varying from hundreds of KHz to a few MHz~\cite{day2022limits, okishev2009stable}. 
In any case, the gate fidelity, obtained by considering the effect of the perturbation on all the Rabi frequencies, is diminished by quantities of the order $10^{-5}$, which is roughly the same loss caused by the transition between quantum gates. As it was argued in the last subsection, this loss in fidelity is negligible compared to the losses due to the presence of noise.

\subsection{An illustrative example}\label{Subsec:example}

As a further application of our GCDD protocol, to illustrate the efficiency of gate transitions in a more complex situation, we show how to protect a quite simple quantum algorithm. To this purpose, we consider the algorithm developed in Ref.~\cite{gedik2015computational}, where it was shown that one can determine the parity in a permutation of three different elements using a single qutrit (it can be extended to determine the parity of $n$ different elements using a single $n$-dimensional quantum system \cite{gedik2015computational}).

If we represent the six possible permutations as
\begin{eqnarray}\label{permutations}
\begin{split}
    &f_1 = \begin{pmatrix} -1 & 0 & 1 \\ -1 & 0 & 1 \end{pmatrix}, \;
    f_4 = \begin{pmatrix} -1 & 0 & 1 \\ 1 & 0 & -1 \end{pmatrix}, \\
    &f_2 = \begin{pmatrix} -1 & 0 & 1 \\ 0 & 1 & -1 \end{pmatrix}, \;
    f_5 = \begin{pmatrix} -1 & 0 & 1 \\ 0 & -1 & 1 \end{pmatrix}, \\
    &f_3 = \begin{pmatrix} -1 & 0 & 1 \\ 1 & -1 & 0 \end{pmatrix}, \;
    f_6 = \begin{pmatrix} -1 & 0 & 1 \\ -1 & 1 & 0 \end{pmatrix}, 
\end{split}
\end{eqnarray}
where the first line indicates the original configuration and the second line the state after the permutation, we have that $f_1$, $f_2$, and $f_3$ are even permutations while $f_4$, $f_5$, and $f_6$ are odd ones. These permutation operations can be easily written as unitary operators. In fact, notice that the unitary $U_X$, defined in Eq.~\eqref{ux}, corresponds to the permutation $f_3$.

It is straightforward to verify that, if we adopt $\ket{0}$ as our initial state, then applying the Hadamard gate $U_H$, defined in Eq.~\eqref{uhadamard}, followed by one of the permutations, followed by $U_H^{\dagger}$, the final state will be, up to a global phase, $\ket{0}$ if the permutation is even and $\ket{1}$ if the permutation is odd. Therefore, a single execution of the circuit involving these three gates can tell us, at the time we measure the final state, the parity of the permutation.

\begin{figure}
	\begin{centering}
		\includegraphics[width=0.48\textwidth]{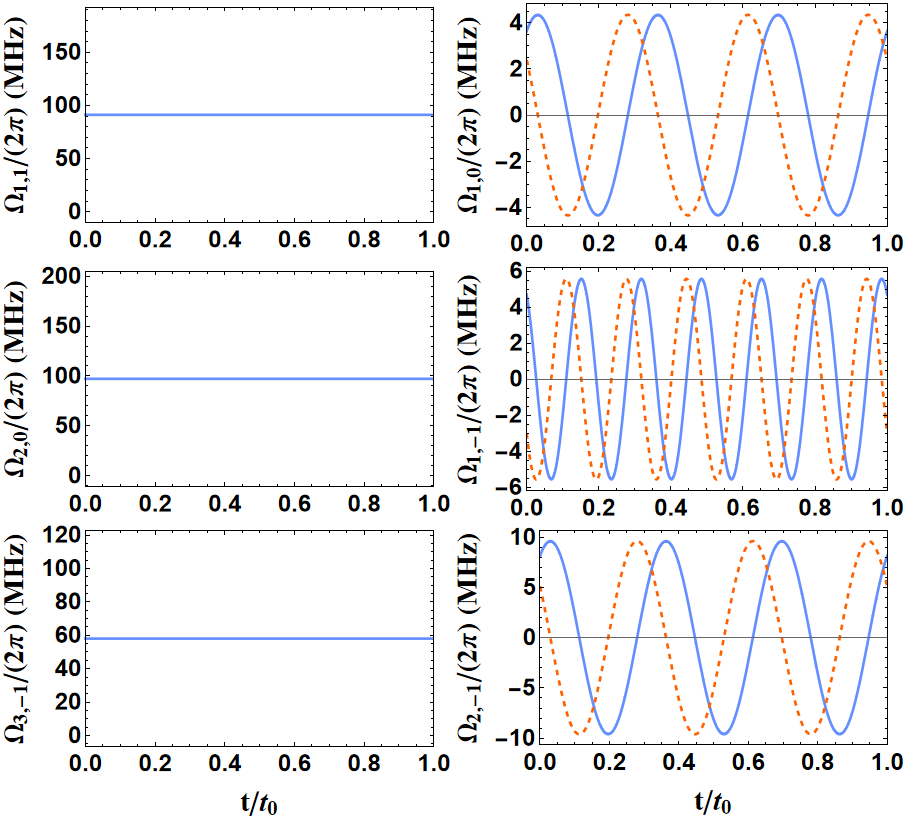}
	\end{centering}
	\caption{\label{f2-Rabis} Same plots and parameters as in Fig.~\ref{Hadamard-Rabis}, but for the $f_2$ permutation gate.}
\end{figure}

\begin{figure}
	\begin{centering}
		\includegraphics[width=0.48\textwidth]{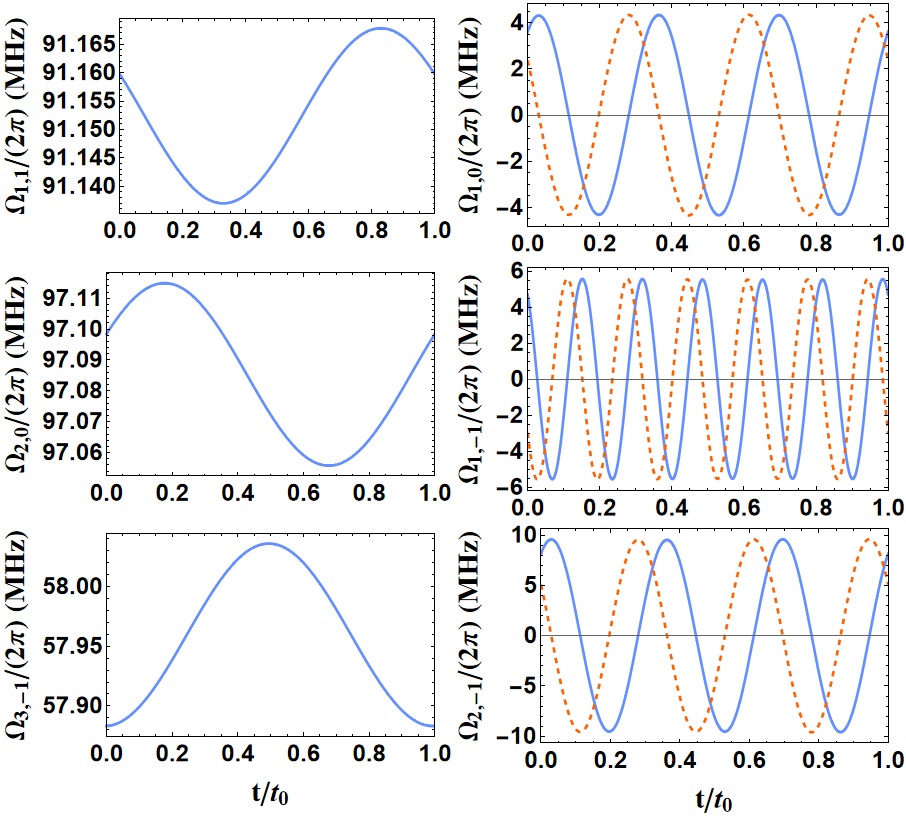}
	\end{centering}
	\caption{\label{f5-Rabis} Same plots and parameters as in Fig.~\ref{Hadamard-Rabis}, but for the $f_5$ permutation gate.}
\end{figure}

In Figs.~\ref{f2-Rabis} and \ref{f5-Rabis} we show the time-dependent Rabi frequencies required to perform the $f_2$ (even) and $f_5$ (odd) unitary operations, respectively.

\begin{figure}
	\begin{centering}
		\includegraphics[width=0.48\textwidth]{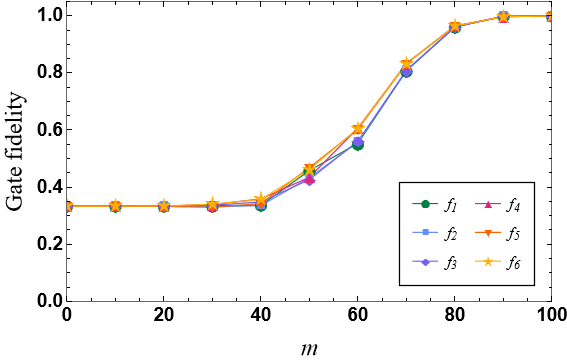}
	\end{centering}
	\caption{\label{superposed}  Impact of the GCDD procedure on the gate fidelity (fidelity at $t=3\tau$) for different values of $m=\tau/t_0$, for the circuit determining the permutation parity for the six different permutations. As required, we set $\ket{0}$ as initial state and apply the three gates in the sequence $U_H^{\dagger} f_k U_H$. The gate fidelity is computed by comparing the obtained state with the expected final state, which is $\ket{0}$ for the even permutations, and $\ket{1}$ for the odd ones. The absence of protection is represented by $m=0$ in the graph. As in Fig.~\ref{Hadamard-ruido}, we use $\tau = 2 \pi \times 10^{-5}$ s, $T = 300$ K, $\lambda_{\text{me}}= 10^{-10}$, and $\omega_c/(2 \pi)= 10^6$ Hz.}
\end{figure}

Using the strategy presented in Sec.~\ref{Subsec:universal} we also simulate the execution of the quantum circuit under the action of both, vacuum and thermal noise, with and without the presence of protection provided by the GCDD method. The simulations are done for the six permutations in Eq.~\eqref{permutations}, and the results for the fidelity of the final measurement, which reveals the parity of the permutation, are all shown (superposed) in Fig.~\ref{superposed}. The gate fidelity is calculated by comparing the density operator obtained at $t=3\tau$ and the expected ideal result, as $\Tr\left\{ \ketbra{j}{j} \rho(3\tau)\right\}$, where $j=0$ for the even permutations and $j=1$ for the odd ones. For all cases we set $\rho(0) = \ketbra{0}{0}$ as required by the protocol.
Notice that for the case where the protection is active with large $m$, the circuit can be executed with almost unitary fidelity for every permutation. For the case where the protection is turned off, the fidelity tends to $1/3$ for all cases (this point is represented by $m=0$ in the graph), corresponding to an evolution to the maximally mixed state.

The results of the simulation with active protection and large $m$ show that, if one of the six permutations $f_k$ is treated as a black box operation inside the circuit, one could infer the parity of $f_k$ with a fidelity arbitrarily close to $1$ by measuring the final state. If we are allowed to use ancilla qudits, it is possible to even guess the permutation $f_k$ instead of just its parity. This could be done by adapting the phase estimation algorithm to qudits~\cite{cao2011quantum}, and applying it right after the described circuit. By measuring the eigenvalue associated with the operation $U_H^{\dagger} f_k U_H$ one could infer the value of $k$, and hence, the permutation.

\section{Conclusions}\label{Sec:Conclusions}

This study is based on the GCDD protection scheme introduced in Ref.~\cite{napolitano2019protecting}, that  provides a method to compute the time-dependent control Hamiltonians required to decouple the system under study from its environment. The major question tackled in this article is whether it is possible to tailor the time-dependent Rabi frequencies required by the GCDD procedure, using the available nowadays technological capabilities. 

We have presented time-dependent Rabi frequencies to decouple an atomic qutrit from both dephasing and amplitude-damping noises occurring together, while also executing intrinsically different quantum operations, including the case of random gates. We notice that the detunings involved in the protection scheme are constant and independent. Within the validity of the approximations of our original calculations~\cite{napolitano2019protecting} we have presented an illustration of how the GCDD control could be realized in the laboratory. We have suggested for the various involved parameters values and orders of magnitudes which might be achievable today in a realistic experiment. Indeed, in the case of manipulation of trapped ions for quantum computing, it is already usual in the laboratory to follow several procedures that, combined, illustrate more technological prowess than the GCDD method requires~\cite{Pino2021}.  We have also considered the protection  of a sequence of quantum gates on the system, showing how the required Rabi frequencies should vary at the transitions. Finally, we have explicitly shown that the protection scheme can be used to implement efficiently a simple quantum algorithm allowing one to determine the parity of the permutations of a set of three elements.

The analysis we present here shows that, at least in principle, it is possible to realize an experiment to test our protection scheme if we are able to emulate the noise effects involved in our model, either by finding a naturally noisy system similar to our setup, or by artificially introducing the same kinds of perturbations.
If such an experiment is accomplished, the principles used in the present model could be translated to other systems with the same logical capabilities and even for qudits with $d>3$. The latter condition could be also realised using atomic qudits based on an extension of our simple qutrit model. It is always important to have thought experiments illustrating how future realistic implementations should work, and it really matters when the illustrative models show possible drawbacks and obstacles one should face when implementing theoretical ideas in the laboratory. We strongly recommend a serious analysis by experimentalists about how to implement a version of such an experiment, at least as a proof-of-principle, to assess possible unforeseen new questions that might further improve the technological aspects of quantum-information processing and quantum computing.

\begin{acknowledgments}
R.d.J.N. and F.F.F. acknowledge support from Funda\c{c}\~{a}o de Amparo \`{a} Pesquisa do
Estado de S\~{a}o Paulo (FAPESP), Projects No. 2018/00796-3 and 2023/04987-6, and also
from the National Institute of Science and Technology for Quantum
Information (CNPq INCT-IQ 465469/2014-0) and the National Council
for Scientific and Technological Development (CNPq). F.F.F. acknowledges support from ONR, Project No. N62909-24-1-2012.
\end{acknowledgments}


%

\end{document}